\documentclass[letterpaper]{article}
\usepackage{aaai16}
\usepackage{times}
\usepackage{helvet}
\usepackage{courier}
\usepackage{graphicx}
\usepackage{balance}
\usepackage{subcaption}
\usepackage{graphics} 
\begin{document}
%
\title{What the Language You Tweet Says About Your Occupation}
\author{Tianran Hu, Haoyuan Xiao, and Jiebo Luo \\
Department of Computer Science \\
University of Rochester \\
Rochester, NY 14623 \\
\{thu, hxiao6, jluo\}@cs.rochester.edu \\
\And Thuy-vy Thi Nguyen \\
Department of Clinical and Social Sciences in Psychology\\
University of Rochester\\
Rochester, NY 14623\\
thuy-vy.t.nguyen@rochester.edu\\
}
\maketitle
\begin{abstract}
\begin{quote}

Many aspects of people's lives are proven to be deeply connected to their jobs. In this paper, we first investigate the distinct characteristics of major occupation categories based on tweets. From multiple social media platforms, we gather several types of user information. From users' LinkedIn webpages, we learn their proficiencies. To overcome the ambiguity of self-reported information, a soft clustering approach is applied to extract occupations from crowd-sourced data. Eight job categories are extracted, including Marketing, Administrator, Start-up, Editor, Software Engineer, Public Relation, Office Clerk, and Designer. Meanwhile, users' posts on Twitter provide cues for understanding their linguistic styles, interests, and personalities. Our results suggest that people of different jobs have unique tendencies in certain language styles and interests. Our results also clearly reveal distinctive levels in terms of Big Five Traits for different jobs. Finally, a classifier is built to predict job types based on the features extracted from tweets. A high accuracy indicates a strong discrimination power of language features for job prediction task.

\end{quote}
\end{abstract}

\section{Introduction}

The recent statistics published by LinkedIn revealed that increasing a number of US employers relies on personality assessments as a way to screen their prospective employees (26\% in 2001 to 57\% in 2013\footnote{https://www.linkedin.com/pulse}). It appears that certain jobs are more likely to attract or fit better with certain people. While this hiring practice has been commonly used by organizations, there has not been any research that directly looks at the connections between different job positions and the people that are in those positions. Indeed, many studies on sociology and psychology suggest that many aspects of a person's lives are deeply connected with their jobs~\cite{lindquist1997influence,strully2009job}. In this study, we looked at the connections between different job positions and the characteristics of the people that hold those positions, including the language that they use, the topics that they are interested in, and their personalty traits. Our findings suggested that jobs are more than just what people do, but also who they are, and how they present themselves to others.


To uncover the connections between jobs and the characteristics of people who hold those jobs, the procedure involves several steps. First, we need to separate people based on their jobs. Categorizing people into job types is particularly challenging because most jobs are less specialized but instead might share some common responsibilities or duties. Therefore, in this step, we rely on the skills that people have to determine the job categories that they are more or less likely to belong to. A soft clustering method is used to assign a normalized weight to each job type for a person based on her skills. 
In the next step, we also need to learn about each person's characteristics, such as her linguistic patterns, interests, personality traits, and so on. Achieving those two steps allow us to compute the Pearson correlation coefficients between people's characteristics and their weights on different job types. To achieve those two steps, we need to first figure out people's skills, and then learn about their characteristics.

Social media posts have been proven an effective information source in gaining knowledge of people~\cite{silva2014you,schwartz2013personality}. Individuals tend to use social media sites as platforms for self-presentation~\cite{schau2003we}. Therefore, we extract rich information about people's characteristics from what they share online. We select Twitter data to support our study, because Twitter is widely and effectively used in user profiling~\cite{nguyen2013old}, and it is relatively easy to collect. For the linguistic features, we apply closed vocabulary and open vocabulary approaches described in~\cite{schwartz2013personality} to ones' tweets. The closed vocabulary approach uses a fixed lexicon to analyse text, while the open vocabulary approach does not limit the vocabulary. In the former, we apply Linguistic Inquiry and Word Count (LIWC) to extract 92 linguistic features from tweets. In the latter, two types of linguistic features are learned: representative words and phrases 
, and topics. Tweets also provide rich information on personalities~\cite{schwartz2013personality}. We apply the IBM Watson Personality Insights service API to compute the personality traits from tweets. Personality is measured by Big Five Traits, a widely examined theory of the five broad dimensions describing human personality~\cite{goldberg1993structure}. The five dimensions are, Openness, Conscientiousness, Extraversion, Agreeableness, and Neuroticism. 

We take advantage of another social media -- LinkedIn -- to collect the job information. 
LinkedIn users share their industries, career experiences, education status, interests, and so on. A skill list is also displayed on a user's LinkedIn page. 
The listed skills are endorsed by the users who knows this person. Each skill in the list is associated with the number of votes this skill receives. Although we can learn LinkedIn users' industries from their profiles, this self-reported information is usually imprecise and ambiguous. For example, we manually check the profiles of all the professors in our department. Some of them report the industry as ``Computer Science'', others fill in ``Higher Education'' or ``Research''. To come up with more precise and accurate job categories, we apply a soft clustering method to people's skills endorsed by others, and use a \textit{group} of skills to describe a job. There are three benefits in doing this: 1) Skill is a concept of finer granularity than industry. 
2) Skills of a person are voted by the people who know the person. The crowdsourced data is more accurate than the information provided by the person him/herself. 3) Given the complex nature of today's jobs, it makes more sense not to restrict a person to only one occupation. Instead, one individual can perform several responsibilities and possess several skills that might possibly qualify him or her for multiple job categories. For example, the job of a  product manger in an IT company could be described with both ``programmer'' and ``manager''.

To match LinkedIn and Twitter accounts of the same people, we collect the users from about.me\footnote{about.me}. It is an integration platform, which allows users to link multiple online identities, such as Twitter, LinkedIn, Facebook, and so on in one profile. We randomly sample about.me users, and retain valid users who register both Twitter and LinkedIn accounts. 

Interesting and significant differences are uncovered among people with different jobs in this study. 
The open vocabulary approach provides more dimensions implying the traits of jobs. For example, office clerks talk more about daily life than people of other jobs. 
According to our results, people with different jobs clearly have different interests. People who are doing public relation and writers show strong interests in politics and social events. 
The divergences of personalities are also found. For example, managers are the most extroverted, while software engineers are the least. 
Base on these observations, we build a classifier to predict jobs using the features extracted from tweets. A high accuracy (80\%+) indicates that adequate personal trait information is hidden in social media posts. 

Our contributions are threefold:
\begin{itemize}
\item We propose an approach to categorize proficiencies by softly grouping skills a person has, using skill information gathered from LinkedIn.
\item We uncover significant and interesting divergences of linguistic patterns, interests, and personalities among people with different jobs.
\item We build up an occupation classifier with high accuracy is built up based on features extracted from Twitter posts. 
\end{itemize}

\section{Related Work}
Studies of sociology and psychology reveal that many aspects of human life, such as health condition, lifestyle, personality, and so on, are deeply connected with their jobs. The effects of work stress on long-term blood pressure is studied in~\cite{lindquist1997influence}. The paper finds that it is the ways of dealing with work stress rather than the stress itself that are significantly related to blood pressure. In~\cite{strully2009job}, the correlation between employment status and health condition is studied. The authors reported that employment status may impact certain health outcomes. 
In addition, sickness absence, the use of alcohol, and anxiety of reorganization are also proven to be related to working life~\cite{voss2004job}. Jobs also influence lifestyles. Payne et al. discussed the different lifestyles of employees in high-strain jobs and low-strain jobs~\cite{payne2002impact}. They found that people with high-strain jobs exercise far less than those with low-strain jobs. The relationship between the Big Five traits of personality and job criteria is investigated in~\cite{salgado1997five,hurtz2000personality}. The findings indicate that Conscientiousness and Neuroticism are valid predictors for job performance. \cite{judge2002five} investigates job satisfaction and personality. The authors found that, job satisfaction is positively correlated with Extraversion, Agreeableness, and Conscientiousness, while it is negatively correlated with Neuroticism. There is also work focusing on specific job types. Perceived social support, job stress, health, and job satisfaction among nurses are studied in~\cite{bradley2002social}. ~\cite{reichel1984job} compared the U.S. hospitality industry employees' with other industries on work attributes, demographics, and class perceptions. The results show that the people of this industry tend to be less satisfied with the job, and take their work as unimportant elements in their self-accomplishments. 

Although much work has been done, previous work on the relationship between human characteristics and jobs focuses on either employment and work status, or a single job type. It is still not clear what divergences exist across various jobs. Due to the inherent limits of transitional data collection methods, previous work of sociology and psychology usually suffers from the problem of small sample size. In our work, a large population is collectied from social media platforms, and we compare multiple human characteristics across jobs.

The psychological meaning of words is well studied in Computer Science and Linguistics. Linguistic Inquiry and Word Count (LIWC), as a computerized text method, is introduced in ~\cite{tausczik2010psychological}. ~\cite{pennebaker1999linguistic} investigates the difference of individual linguistic styles. This work reports the significant difference across their language patterns, and proves the effectiveness of LIWC. Based on the linguistic features, ~\cite{mairesse2007using} introduce an approach to recognizing personalities from conversation. The boom of social media attracts a lot of research that are based on this new data source. It is shown that personalities can be recognized using people's social media network structures~\cite{staiano2012friends}, profiles~\cite{quercia2011our}, and contents of posts~\cite{qiu2012you,golbeck2011predicting}. ~\cite{vinciarelli2014survey} provide a nice survey on computing personality from social media data. In this paper, we follow the approaches described in~\cite{schwartz2013personality} to extract linguistic patterns. In their work, Schwartz et al. propose two approaches to learning people's language styles from social media texts. They report significant differences in language styles across several features including genders, ages, and personalities.

\section{Data Collection and Preprocessing}
\subsection{User Collection}
We use about.me search API\footnote{about.me/developer/api/docs/} to collect users. The input of this API is a name, and it returns the information of at most 100 users who has the same or a similar name. We gather the 1,000 popular male and female first names in the U.S, and feed these 2,000 names to the API. The API returns 150K user profiles. We first select the users who have both LinkedIn and Twitter accounts. Following their Twitter accounts, we download these users' 3,000 most recent tweets using Twitter search API\footnote{dev.twitter.com/overview/documentation}. We remove the users who do not have enough English tweets (less than 2,000), to guarantee the significance of our results. We then collect the job information through LinkedIn links. We also remove the people who does not have a skill list on the LinkedIn page, because we cannot categorize their jobs for them. Eventually, we end up with 9,800 users in total. 

\begin{table*}
\centering
\begin{tabular}{ | l || c | c | c | c | c | }
\hline
Job Name & First Skill & Second Skill & Third Skill & Fourth Skill & Fifth Skill\\ \hline \hline

1. Marketing & Digital MKTG & Social Media MKTG & Online MKTG & Digital Strategy  & Advertising\\ \hline
2. Administrator & Public Speaking  & Leadership & Fundraising & Event Planning & Coaching\\ \hline
3. Start-up & Start-ups & Entrepreneurship & Strategy & Business DEV & Management\\ \hline
4. Editer\&Writer & Blogging & Editing & Journalism  & Copy Editing & Storytelling\\ \hline
5. Software Engr & MySQL & CSS & JavaScript & PHP & jQuery\\ \hline
6. Office Clerk & Microsoft Office & Microsoft Excel & PowerPoint & Microsoft Word & Customer Service\\ \hline
7. Public Relation & Public Relations & Media Relations & Press Releases & Strategic COMM & Corporate COMM\\ \hline
8. Designer & Graphic Design & Web Design & Photography & Illustrator & Photoshop\\ \hline
\end{tabular}
\caption{Extracted job categories using LDA. We list 5 most weighted skills of each job, and manually label them according to the skills. ``Engr'', ``COMM'', ``MKTG'', and ``DEV'' are short for Engineer, Communications, Marketing, and Development, respectively.}~\label{tab:jobs}
\vspace{-2em}
\end{table*}

\subsection{Tweets Cleaning and Phrase Selection}
Twokenizer\footnote{http://www.cs.cmu.edu/~ark/TweetNLP} is an NLP tool designed for tweets especially. It detects abbreviations or slang (\textit{b4} for \textit{before}, \textit{fb} for \textit{Facebook}), misspellings or spelling variants (\textit{fir} for preposition \textit{for}), and emoticons (\textit{:)}, \textit{\textless3}) in tweets. However, in practice Twokenizer usually fails in extracting Twitter official emojis\footnote{http://apps.timwhitlock.info/emoji/tables}, because these emojis concatenate with other words in many cases. Therefore, we apply Twokenizer to separate words from tweets, and we also use a fixed emoji list to detect all the official emojis. Terms that are too popular (used by more than 95\% users) and too unpopular (used by less than 10\% users) are removed. 

We apply the same thresholds to phrases (2-grams and 3-grams). To avoid extracting simple word combinations instead of meaningful phrases, we apply Point-wise Mutual Information (PMI) to distinguish these two~\cite{schwartz2013personality}. PMI measures how more informative it is to take a phrase as a whole compared with taking it as separate words. It is formally defined as follows:
\begin{equation}pmi(phrase)=
\log\frac{P(phrase)}{\Pi_{t\in phrase}P(t)}\end{equation} where $t$ indicates the terms in the phrase, and $P(phrase)$ is the probability of observing the phrase. We filter out all phrases with PMI value lower than 2* length, where length is the number of words in this phrase, and we consider these phrases as uninformative.

However, it is infeasible to compute PMI for every phrase, due to the large size of candidate set. We exploit the Apriori property of phrases~\cite{agrawal1996fast} to reduce the candidate set. The apriori property states that if a set is not frequent, then all the super sets are not frequent neither. In our case, for example, if ``love you'' is used by less than 10\% of the users, then ``I love you'' has to be used by less than 10\% of the users. In practice, we first obtain all the 1-gram terms using Twokenizer and emoji list, and remove the 1-gram terms that are used by less than 10\% of the users. We then combine 1-gram terms pairwise to generate 2-gram phrases. The same steps are applied in generating 3-gram candidate set from 2-gram phrases. At last, we remove words and phrases that are used by more than 95\% of the users from the candidate sets, and apply PMI to remove the uninformative phrases. The preprocessing leaves us 18,082 words, 7,852 2-gram, and 10,903 3-gram phrases.

\section{Job Categorization}
Most LinkedIn users report their industries in the profiles, such as ``Marketing'', ``Public Relation'', and ``Web Design''. However, these industry tags are not precise enough; and some of them are even ambiguous. We observed many cases where people with different industry tags actually work in the same position. For example, a computer science professor could report his/her industry as ``Computer Science'', ``Higher education'' or ``Research''. To learn the real job a person is doing, we utilize the skills a person has. A person's skills are endorsed by people who connect with this person, and are listed in one's linkedIn page. Besides a list of endorsed skills, the number of endorsements of each skill is also available. 

We vectorize a person's skill list into a fixed order vector. Each dimension represents a skill, and the values of elements in a vector are the votes this person receives on the corresponding skills. We feed the skill vectors to Latent Dirichlet Allocation (LDA) model to generate a soft clustering of people. Two matrices are learned using LDA. The topic matrix describes the distribution of skills in each cluster. In other words, clusters are linear combinations of skills. These clusters are used as definitions of job in this study. The document matrix describes the people's weights on each job. Therefore, each person is not limited to one job. These assigned weights are used to calculate the correlations between jobs and linguistic patterns, as well as personality traits. There are three benefits from categorizing jobs in this way. First, comparing with a broad industry, it is more precise to describe proficiency with a set of skills. Second, as skills of a person are voted by others, the crowdsourced data is more accurate than the self-reported information. Third, soft clustering solves the ambiguity in defining one's job. Taking the example we used in introduction, now a product manger in an IT company can be defined as 30\% ``programmer'' and 70\% ``manager''.

The number of clusters (topics) $k$ is a crucial parameter in LDA setting. We use perplexity to determine how many topics are needed. Perplexity measures how good a language model predicts unseen documents, and decreases monotonically as the number of topics increases~\cite{blei2003latent}. We use 10-cross validation to test $k$ from 2 to 100. Average perplexity drops fast as $k$ increases from 2 to 20, and keeps almost unchanged as $k$ goes above 20. Therefore, we select number of cluster as 20. To guarantee that each job covers a relatively large population, we 
remove the jobs that contains less than 300 users. We end up with 8 jobs after these processes. In Table~\ref{tab:jobs}, we list the 5 most heavily-weighted skills for each job category and also manually assign a label for each job category. The results are consistent with common sense. Take the fifth job ``Software Engineer'' as an example, the skills associated with this job are mostly programming languages (\textit{JavaScript}), and tools (\textit{MySQL}). The remaining 7 jobs are Marketing, Administrator, Star-up, Editor\&Writer, Office Clerk, Public Relation, and Designer.

Please note that we do not aim at finding out all the job types, but focus on the divergences among people with different jobs. Therefore, we did not try to extract an exhaustive or complete list of jobs in this study.

\begin{figure}[!htbp]
\centering
\includegraphics[width=0.9 \columnwidth]{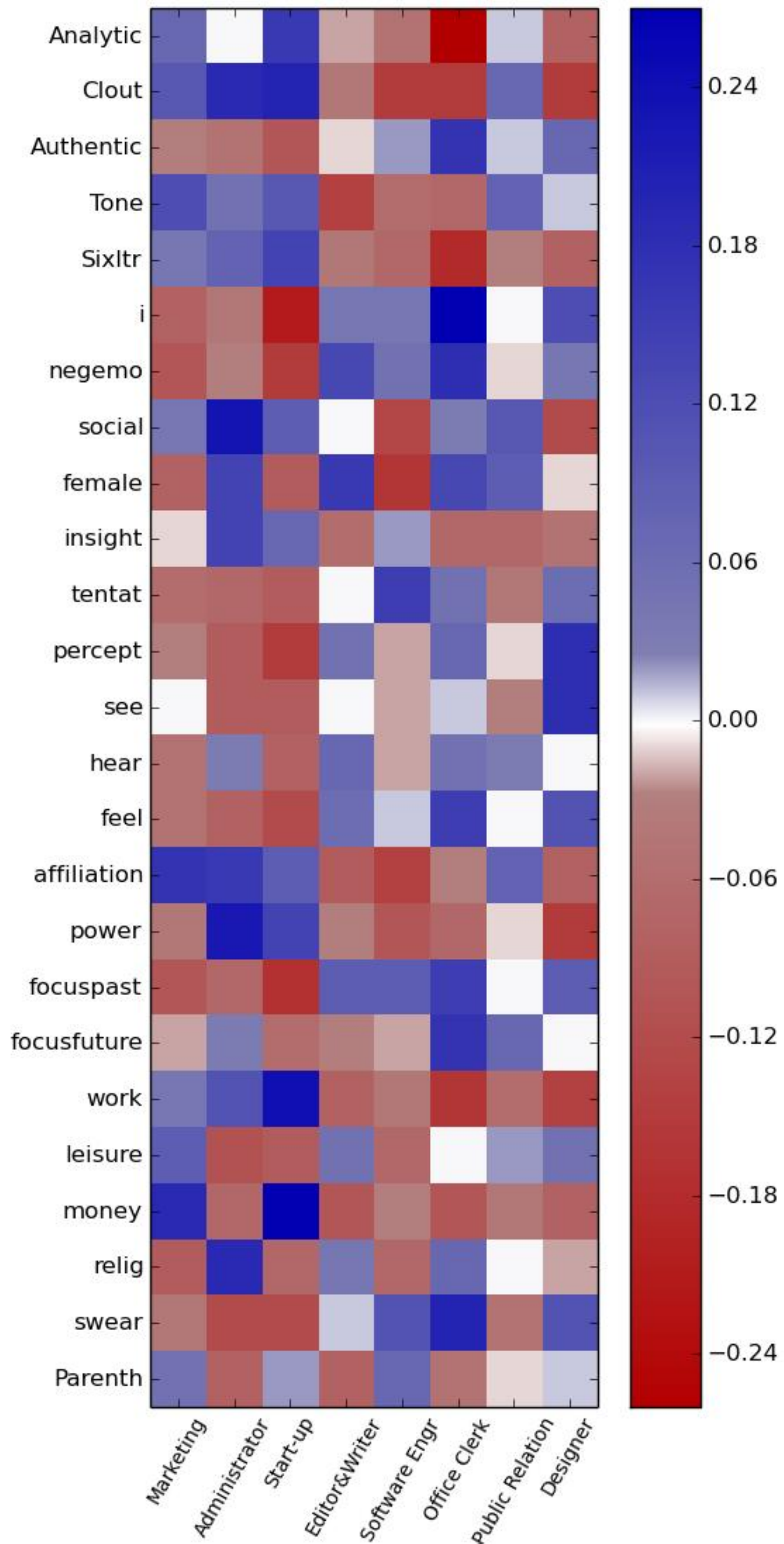}
\caption{25 LIWC features, and their correlations with each occupation.}~\label{fig:LIWC_job}
\end{figure}

\section{Closed Vocabulary Approach and Results}
We first used a fixed lexicon (LIWC) to analyse the linguistic patterns of different jobs. Since the vocabulary is fixed in the study, this approach is called closed vocabulary approach in~\cite{schwartz2013personality}. Linguistic Inquiry and Word Count (LIWC) is a computer program for language analysis~\cite{pennebaker1999linguistic}. LIWC2015 computes a 92-Dimensional vector for a document based on a given dictionary. The dimensions include: Sixltr (percentage of words that longer than 6 letters), Anger (percentage of words that indicates anger), and so on.

We combine all the tweets of a person as a single document, and then calculate the LIWC features. Pearson Correlation Coefficients are computed to uncover the unique language traits of different jobs. Due to the length limit, in Figure~\ref{fig:LIWC_job} we list 25 LIWC features, and their correlations with each job. The p-value of these correlations are all smaller than $10^{-4}$. Significant and interesting correlations are observed. We summarize them as follows (the terms in brackets are the feature names of LIWC outputs):

\begin{itemize}
\item Marketing people talk a lot about money and affiliation (\textit{money}, \textit{affiliation}), and they mention more about leisure (\textit{leisure}). Their emotional tone scale (\textit{tone}) is high. A higher emotional tone scale indicates more lively than other people
. Meanwhile, they talk less about religions (\textit{relig}), and show fewer negative emotions (\textit{negemo}).
\item For the administrators, they prefer to use first-person plural, and mention more about power, insight and society. On the other hand, this group of people use fewer swear words (\textit{swear}) and parentheses (\textit{Parenth}) comparing with people of other jobs.
\item People in start-ups post a lot of contents about money and work (\textit{work}) in their tweets, while they do not use first-person singular much, and they do not express anger or negative emotion much.
\item As to the editors and writers, they use many words about genders (\textit{female}, \textit{male}). Different from other jobs, people with this job mention more about negative emotions. 
\item Software Engineers' posts are more tentative (\textit{tentat}), and the usage of swear words is positively correlated to the job. Not surprisingly, programmers mention less about female and social. Likewise, Programmers' power-awareness (\textit{Clout}, \textit{power}) is relatively low.
\item Office clerks mention a lot about themselves (\textit{i}) in their tweets, and the usages of negations (\textit{negat}), negative emotions are higher. People doing this job post fewer analytic tweets (\textit{Analytic}) and long words (\textit{Sixltr}), and they do not talk much about work.
\item People who are doing public relation tend to be more future-focused (\textit{focusfuture}). In other words, they like to use words like \textit{will}, \textit{may}, and \textit{soon} in their tweets. Same as marketing people, this group of people's emotional tone scale is relatively high.
\item Designers like to describe what they see, hear, and feel (\textit{seen}, \textit{percept}) in tweets. Similar to programmers, they have low power-awareness (\textit{Clout}, \textit{power}).
\end{itemize}

\section{Open Vocabulary Approach}
In addition to using a fixed lexicon to analyse the linguistic patterns of different jobs, in this section we discuss an approach leveraging all the words in ones' posts. This approach is called the Open Vocabulary Approach in~\cite{schwartz2013personality}. We first count the words and phrases in a person's tweet, then use TF-IDF to weight all the terms based on their frequencies. Pearson Correlation Coefficients are computed between the people's TF-IDF weights of each term and their weights on each job. 
The positively correlated terms to a job are used more by the people with the job. In other words, these terms are positively distinguishing to this job. Oppositely, negatively correlated terms to a job are used less often by the people with the job, they are negatively distinguishing to this job. After this, we aggregate each person's tweets into a single document, and feed the documents to LDA to extract 2,000 topics. We then calculate the correlation between people's weights on the 2,000 topics and their weights on the 8 jobs. By doing this, we find the positively and negatively distinguishing topics to a job. Because topics extracted from tweets are meaningful combinations of words, we can learn more about the interests and focusings across people with different jobs.


\begin{figure*}[!htbp]
 
\begin{subfigure}{0.49\textwidth}
\includegraphics[width=0.9\linewidth, height=5cm]{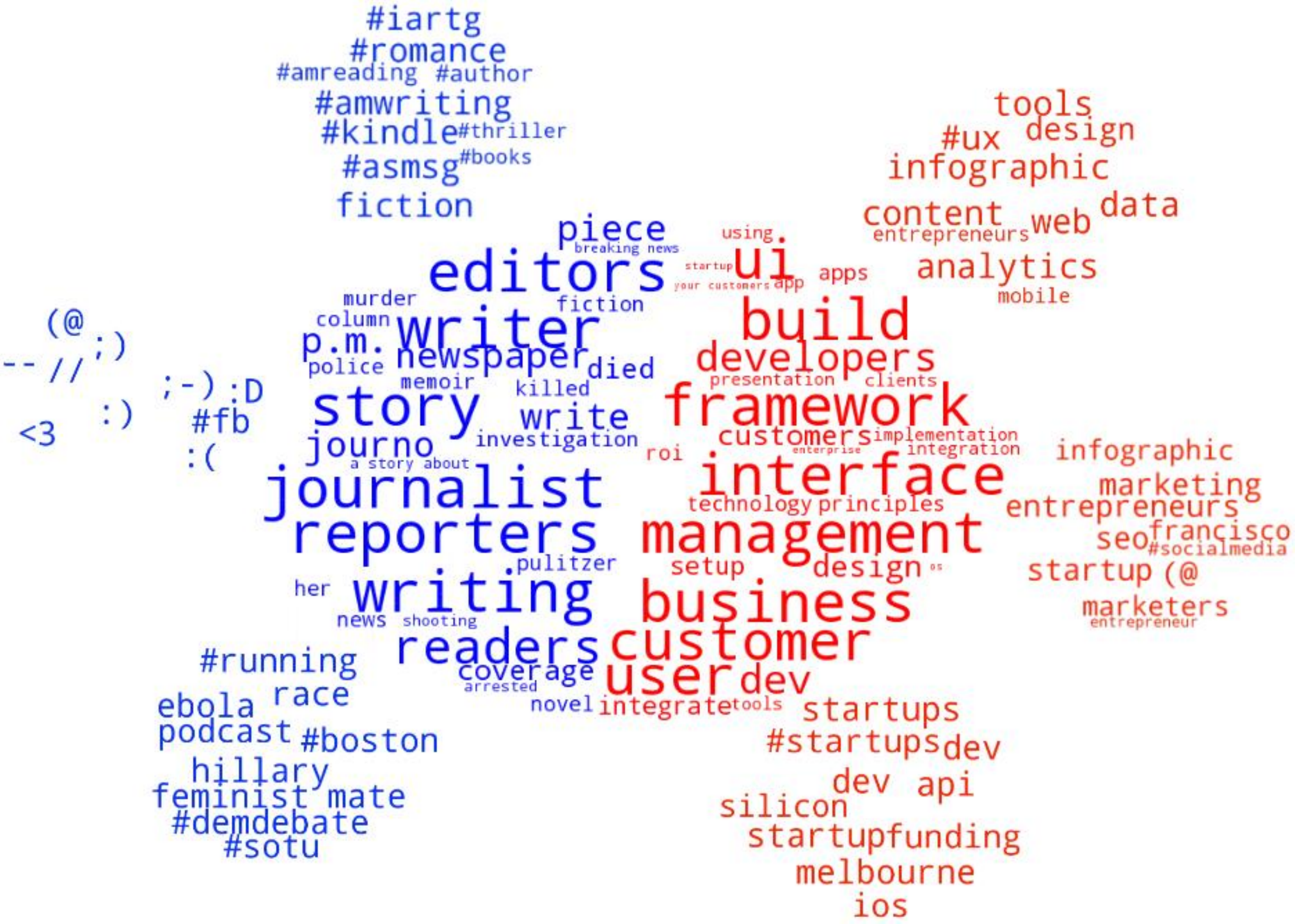} 
\caption{Word cloud for Editors\&Writers}
\label{fig:subim1}
\end{subfigure}
\begin{subfigure}{0.49\textwidth}
\includegraphics[width=0.9\linewidth, height=5cm]{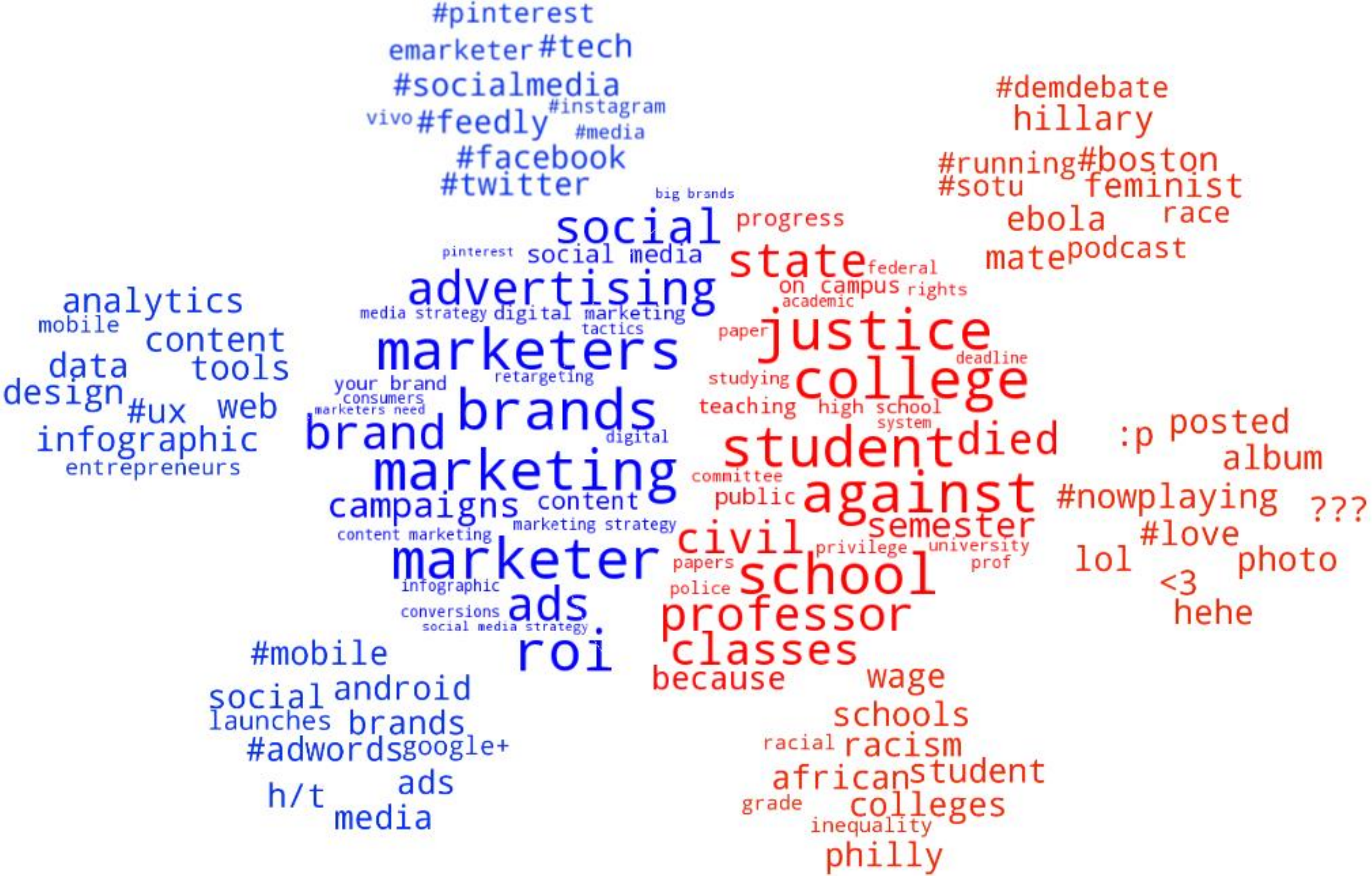}
\caption{Word cloud for Marketing}
\label{fig:subim2}
\end{subfigure}
 
\caption{A word cloud illustration of highly correlated words, phrase, and topics of two occupations: Marketing and Editors\&Writers. Positively correlated ones are plotted in blue (left), and the negative ones are in red (right). The big circle in the middle contains the highly correlated words and phrases. Small surrounding circles represent topics, with each circle being a topic. We select top 200 positively and negatively scores words and phrase, and for each topic we select the top 10 prevalent words.}
\label{fig:word_cloud}
\vspace{-1em}
\end{figure*}

\subsection{Results of Open Vocabulary Approach}
We find interesting differences across jobs. Each job has its unique preference to words, phrase, and topics, suggesting the divergences on working content, interests, and characteristics of people with different jobs. Some topics are highly positively correlated to one job, while highly negatively correlated to others. We plot these distinctive terms and topics in the form of word cloud (Figure~\ref{fig:word_cloud}), due to the length limit, we only plot two jobs. We summarize our results as follows.

\subsubsection{Marketing}
The positively correlated terms and topics indicate that the most distinguishing contents to marketing people are about the methods, platforms, and strategies of marketing. Meanwhile, this group of people have less interests in politics, policy, and education. 

The most positively correlated word is \textit{marketers}, indicating the occupation of this group of people. Since social media is growing into a more and more important marketing platform, the correlation of \textit{social media} is quite high. Others words and phrases, such as \textit{your brand}, \textit{targeting}, \textit{campaigns}, \textit{marketing strategy}, and \textit{lead generation}, are also extracted, implying the unique tweet contents of marketing people. To our surprise, many most negatively correlated words and phrases to this job are about education, such as \textit{paper}, \textit{on campus}, \textit{professor}, and \textit{semester}. Meanwhile, some legal words, such as \textit{justice}, \textit{evidence}, \textit{civil}, \textit{federal}, and \textit{violence}, are also highly negatively correlated to the people who have this job.

The most positively correlated topic is about mobile and mobile advertising to this job (\textit{adroid}, \textit{google}, \textit{adwords}, \textit{brands}, and \textit{ad}). The following two topics are about data analysis (\textit{analytics}, \textit{infographics}, \textit{data}), and social media platforms (\textit{twitter}, \textit{pintrest}, \textit{facebook}, \textit{instagram}), respectively. The most negatively correlated topic is formed by words about school and politics (\textit{inequality}, \textit{racial}, \textit{racism}, \textit{college}, \textit{students}), which aligns with the negatively correlated words and phrases. The second most negatively related topic is about politics (\textit{hillary}, \textit{demdebate}\footnote{meaning: Democratic Party Presidential Debates}, \textit{\#sotu}\footnote{meaning: State of the Union}) and social events (\textit{ebola}, \textit{feminist}). 

\subsubsection{Administrator}
Administrators talk more about leading, education, and religion in their tweets, while they do not express negative emotions or use negations much. Comparing with other jobs, these group of people are more careless about techniques.  

Administrators' tweets are more invigorated. They use many didactic words and phrases like \textit{make a difference}, \textit{courage}, and \textit{honor}. Vocabularies related to leading are also frequently used. For example, words and phrase, such as \textit{leaders}, \textit{we must}, and \textit{leadership} have high positive correlation scores. Meanwhile, administrators prefer to use phrases related to unity such as \textit{we need to}, \textit{we must}, \textit{we are}, \textit{how do we}. This agrees with our finding using closed vocabulary approach about their preference for useing first-person plurals. There are many religious vocabularies highly correlated to them too, such as \textit{blessing}, \textit{pray}, and \textit{god's}. Different from marketing people, education vocabularies (\textit{students}, \textit{youth}, \textit{college}) are positively correlated to this group of people. A large percentage of most negatively correlated words to managers is related to techniques, such as \textit{app}, \textit{version}, \textit{iphone}, \textit{ui}, and \textit{interface}.

The most positively correlated topic to this job is about education and politics (\textit{inequality}, \textit{racial}, \textit{racism}, \textit{college}, \textit{students}). Interestingly, this topic is among the most negatively correlated topics to marketing people. A topic about religion (\textit{worship}, \textit{jesus}, \textit{bible}, \textit{lord}) also has high positive correlation with this job. This is consistent with the fact that religious vocabularies are positively correlated. The third most positively correlated topic is about education (\textit{\#edchat}, \textit{\#edtech}, \textit{\#education}, \textit{tearchers}, \textit{\#edtechchat}). The most prevalent terms in this topic are usually attached with hash-tag, indicating their interests on related discussion on Twitter. The most negatively correlated topic is formed by  negation terms (\textit{dont}, \textit{cant}, \textit{:d}), and negative emotion words, such as \textit{jealous} and \textit{bloody}. The second topic is about popular mobile device (\textit{ios}, \textit{apple's}, \textit{iphone}), and sports (\textit{nfl}, \textit{nike}, \textit{sports}). The insulativity of managers from programming skills is revealed by the third most negatively correlated topic, consisting mainly of programming words (\textit{php}, \textit{js}, \textit{api}, \textit{github}, \textit{jquery}).

\subsubsection{Start-up} 
Start-ups people share some likes and dislikes with marketing people and mangers. Avoidance of using self-denial expression is one of their unique features.

Almost all the highly positively correlated words and phrases, such as \textit{founders}, \textit{investors}, \textit{growth}, \textit{valuation}, and \textit{companies}, are about running company, investment, and business. Not surprisingly, \textit{silicon} is also one of the most positively correlated words. Like administrators, start-up people dislike to use negating words (\textit{can't}, \textit{dont}), but they especially dislike self-denial. \textit{i can't} is the most negatively correlated phrase to this group of people. This dislike of self-denial is also reflected by the high negative correlation score of \textit{i don't know} and \textit{i didn't}. \textit{she} and \textit{her} also do not appear much in their tweets.

The most positively scored topic contains words like \textit{bitcoin}, \textit{tesla}, \textit{crowdfunding}, \textit{startups}, \textit{inverstors}, and so on. The following two topics are also among the top positively correlated topics of marketing people. These two topics are about data analysis (\textit{analytics}, \textit{infographics}, \textit{data}), and mobile advertising (\textit{adroid}, \textit{google}, \textit{adwords}, \textit{brands}, and \textit{ad}). The overlapping indicates the similar concerns of people with these two jobs. Same as administrators, the most negatively scored topic of start-ups people is the one formed by neglecting terms (\textit{dont}, \textit{cant}, \textit{:d}), and negative emotion words(\textit{jealous}, \textit{bloody}). The following two negative topics mainly consist of emoticons, such as \textit{;)}, \textit{;-)}, and \textit{:(}. 

\subsubsection{Editor \& Writer}
Editors and writers show more interest in politics, as well as in social events. They also talk more about reading and books. Comparing with the jobs we mention above, this people with this job use more emoticons. 

The four most positively correlated words to editors are \textit{editors}, \textit{journalist}, \textit{writer}, and \textit{reporter}, clearly implying the occupation of these people. In addtition to, the words closely related to this job, such as \textit{headline}, \textit{pulitzer}, and \textit{newspaper}. 
We also observed many words related to social events like \textit{murder}, \textit{investigation}, and \textit{police}. 
The two most negatively correlated type of words are: techniques words (\textit{interface}, \textit{setup}, \textit{framework}), and businesses words (\textit{management}, \textit{customer}, \textit{business}). 

The topic that attracts the most attention from this group is the one about politics (\textit{hillary}, \textit{demdebate}, \textit{\#sotu}) and social events (\textit{ebola}, \textit{feminist}). Please note that this topic is also the one that attracts the least attention from marketing people. Editors also would like to talk about reading. The second most positively scored topic is about \textit{\#romance}, \textit{\#thriller}, \textit{\#books}, and \textit{\#kindle}, followed by the third emoticon topics (\textit{;)}, \textit{;-)}, \textit{- -}). The most unpopular topic to editors is about analytics and techniques, words like \textit{data}, \textit{analytics}, and \textit{mobile} are among the representatives of this topic.

\subsubsection{Software Engineer}
Terms and topics indicate that software engineers use way more technique terms than others, while they tend to talk less about females and social life. 

Software engineers mention a lot about techniques and coding (\textit{web}, \textit{ui}, \textit{code}, \textit{plugin}). In fact, we could not find a word or phrase that is not about techniques among the top 200 most positively correlated. The most negatively correlated word to software engineers is \textit{summer}, followed by \textit{girl} and \textit{her}. \textit{relationship} is also a less often used word by them. They also express less interest on excitements and celebrations. For example, \textit{love this!}, \textit{so excited}, \textit{sunday}, and \textit{gift} are all highly negatively scored. Moreover, their expressions of compliment (\textit{thank you}, \textit{cutest}, \textit{so proud of}) appear less than people with other jobs. 

The two most positively correlated topics, with no surprise, are both about programming techniques. The representative words of them are mainly about programming languages (\textit{php}, \textit{java}, \textit{python}), and tools (\textit{github}, \textit{photoshop}, \textit{api}). The last topic among the top three positive topic is about apple products (\textit{ios}, \textit{apple}, \textit{iphone}) and sports (\textit{nike}, \textit{sports}, \textit{nfl}). The most negatively correlated topic to this job is mixed by three types of words: marketing vocabularies (\textit{marketing}, \textit{advertising}, \textit{\#marketing}), family vocabularies (\textit{husband}, \textit{dinner}, \textit{family}), and vocabularies of praising (\textit{fabulous}, \textit{angeles}). The following two topics are formed by Twitter official emojis (please refer to the supplementary material for the illustration) and regular emoticons (\textit{\textless3}, \textit{;)}), respectively. 

\subsubsection{Office Clerk}
The most positively distinguishing words and phrases of people with other jobs are related to their job contents, such as vocabulary of programming languages to software engineers, and vocabulary of reading and writing to editors. However, office clerks' most positively correlated words have nothing to do with their job. They emphasize life and family in their posts instead, and express strong self-awareness. They tend to use more emojis and emoticons, and their negative emotions are stronger than people with other jobs. Negatively scored topics indicate that office clerks show less interest on business, and data analysis. 

The most positively scored term to them is \textit{my life}. Other phrases related to daily life such as \textit{woke up}, \textit{fall asleep}, and \textit{my hair} are also highly scored. Words and phrases of self-expression (\textit{i just want}, \textit{i wish i}, \textit{i hate}) appear a lot in office clerk's tweets. It is interesting that a lot of self-expression portrayed among office clerks are emotional. They mainly express lack of motivation, negative emotions, unfulfilled wants and wishes. Words such as \textit{semester}, \textit{studying}, and \textit{homework} get high positive scores, implying this group of people's strong interest in education. They have a mixed negatively correlated vocabulary. The less usage of \textit{the future of} aligns with the low future-focused score we calculated using the closed vocabulary approach. Technique vocabularies (\textit{web}, \textit{online}, \textit{app}) also appear less in their tweets. Moreover, their posts are less motivated. The terms like \textit{interesting}, \textit{creating}, \textit{great} all have high negative scores.

Contrary to software engineers, this group of people is most positively related to the topic consisting of emojis. Moreover, unlike administrators and star-ups people, the topic formed by neglecting terms (\textit{dont}, \textit{cant}, \textit{:d}), and negative emotion words (\textit{jealous} and \textit{bloody}) is the second most positively scored topic. 
The third most popular topic of this group is about entertainments (\textit{\#nowplaying}, \textit{photo}, \textit{:p}, \textit{album}). As to the negatively correlated topics, the top three are about analytics, mobile advertising, and business, respectively. These topics are positively scored topics to administrators, marketing people, and people from start-ups.

\subsubsection{Public Relation}
People doing public relation show more interest in politics and social events. They also have strong sense of time, according to the high usage of temporal words in their tweets.

The five most positively distinguishing words and phrases to this occupation are \textit{pr}, \textit{\#pr}, \textit{pr pos}, \textit{public relations}, and \textit{press releases}, indicating the job content clearly. Not surprisingly, words related to social events, such as \textit{anniversary}, \textit{super bowl}, and \textit{crisis} are more frequently used by them. They particularly prefer temporal words and phrases, such as \textit{a.m.}, \textit{p.m.}, \textit{of the year}, \textit{monday}. This is probably because time is a crucial factor in public relation. The high positive score of exclamation mark (\textit{!}) implies relatively strong tone of their tweets. This group of people barely mention words about techniques. The top 200 negatively scored words of this job have a large overlapping with the top 200 most positively scored words of software engineers. 

The two most positively correlated topics to this group people are formed by politics and social events vocabularies (\textit{hillary}, \textit{demdebate}, \textit{\#boston}, \textit{ebola}), and emojis, respectively. The third topic is about social media platforms (\textit{\#socialmedia}, \textit{\#facebook}, \textit{\#twitter}, \textit{\#pinterest}, \textit{\#instagram}), indicating the application of social medias in the field of of public relation. The most negatively scored topic is about programming languages and tools, which agrees with the negatively scored words and phrases. 

\begin{figure*}[!htbp]
\centering
\includegraphics[width= \textwidth]{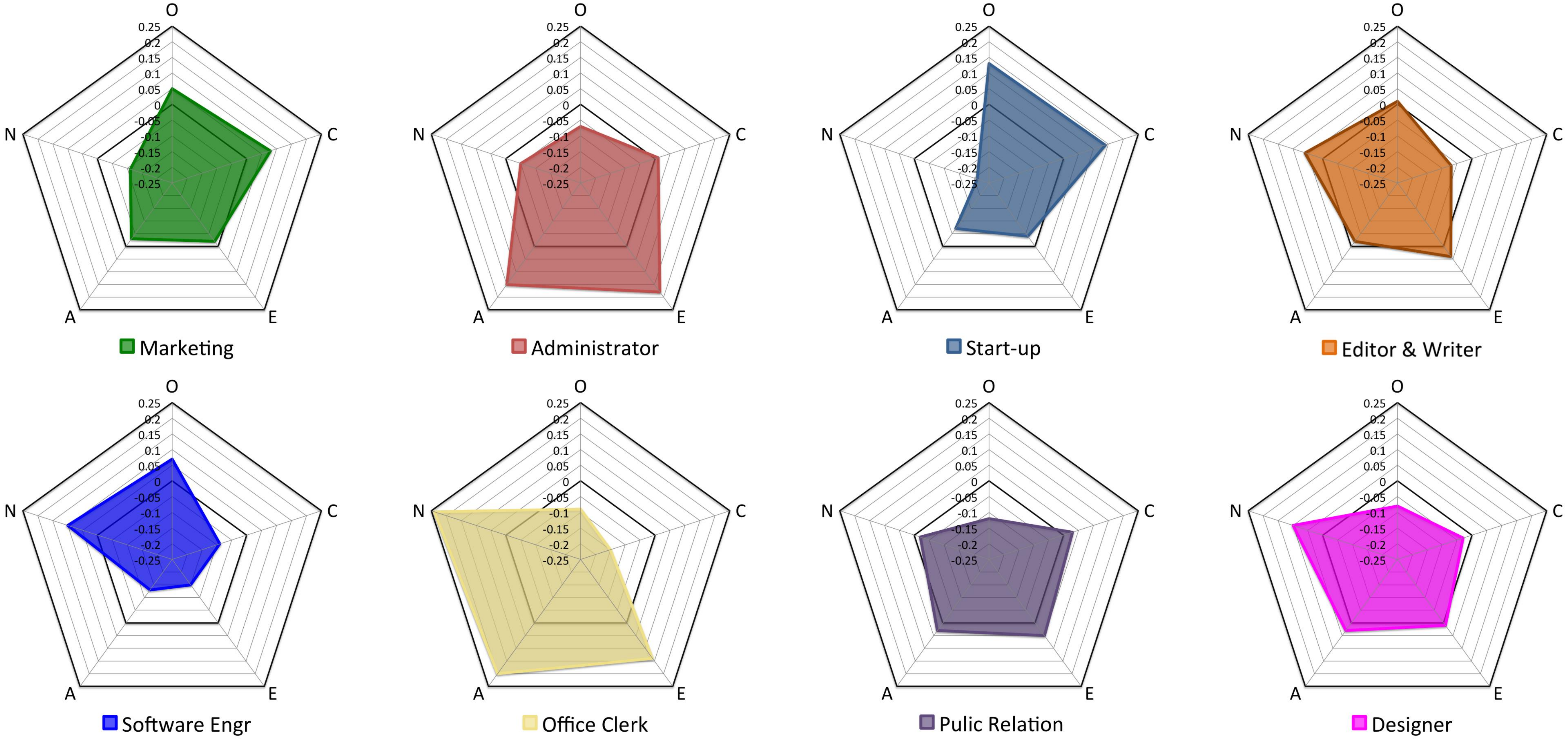}
\caption{Radar plots of Pearson correlation coefficients between personality traits and each job, where O, C, E, A, N stands for Openness, Conscientiousness, Extraversion, Agreeableness, and Neuroticism, respectively. All the correlations are significant with a p-value lower than $10^{-4}$.}~\label{fig:personality}
\vspace{-2em}
\end{figure*}

\subsubsection{Designer}
Designers use more visual-related words and compliment words, while they show less interest in business. They also express special interest in New York City. 

The three most positively correlated words to designers are \textit{illustrator}, \textit{designer}, and \textit{graphic}. Colors (\textit{blue}, \textit{red}, \textit{black and white}) and words related to graphic designing (\textit{font}, \textit{logo}, \textit{icon}) are more frequently mentioned by them. The name of a community where designers and photographers share their works, \textit{behance}, is also positively scored. Compliment words like \textit{cute}, \textit{nice}, \textit{sweet} are more frequently used by designers than people with other jobs. According to the most negatively correlated words, designers have less interest in business (\textit{benefits}, \textit{inverstment}, \textit{strategies}), and companies (\textit{ceo}, \textit{colleagues}, \textit{leadship}).

The most concerned topic to designers is about their job contents. The most representative words of this topic are \textit{typography}, \textit{fonts}, \textit{lettering}. The second topic is mixed with emoticons, and social media platform names. In the third most positively scored topic, besides some emoticons, expressions about New York City (\textit{nyc}, \textit{\#ny}, \textit{ny}) are also highly weighted. We believe this is because the city has special meaning to designers. As to the negatively scored topics, the tops three are about politics and social event (\textit{hillary}, \textit{demdebate}, \textit{ebola}, \textit{\#sotu}, \textit{feminist}), marketing (\textit{seo}\footnote{meaning: Search Engine Optimization}, \textit{\#marketing}, \textit{\#contentmarketing}), and business (\textit{startups}, \textit{league}, \textit{silicon}), respectively. 


It is effective to learn personality from language styles~\cite{mairesse2007using}. Illuminated by the difference of language styles existing across jobs, we expect to observe distinctive personality traits. In the next section, we focus on the divergences of personality of people with different jobs. 


\section{Personality Analysis}

In this section, we study the personality features of the people of different jobs. Previous work shows that people's personalities can be calculated from their tweets~\cite{qiu2012you}. We apply the IBM Watson Personality Insights service API\footnote{www.ibm.com/smarterplanet/us/en/ibmwatson/} to compute the personality traits from ones' tweets. The input is all the tweets of a person, and the service analyzes the linguistic features to infer personality, including Big Five, Needs, and Values. Due to the length limit, we only discuss the Big Five Personality Traits in this paper. Big Five is a widely examined theory of five broad dimensions describing human personality~\cite{goldberg1993structure}. The five dimensions are:
\begin{itemize}
\item \textit{Openness}: measures how open a person is to unusual ideas, imagination, curiosity, and variety of experience. In other words, a higher openness indicates a higher acceptance to new things and changes.
\item \textit{Conscientiousness}: reveals a person's self-discipline. People of higher conscientiousness tend to act in an organized and thoughtful way.
\item \textit{Extraversion}: indicates the extent to which a person prefers or enjoys being in social situations or interactions with the outside, and have company of others. 
\item \textit{Agreeableness}: reflects if a person feels comfortable about compromising. Higher agreeableness implies being more cooperative toward others.
\item \textit{Neuroticism}: measures the instability of a person's emotions. It is usually easier for a person of higher neuroticism to experience negative emotions.
\end{itemize}
Our results reveal clear distinctive levels of the Big Five Traits of different jobs. In Figure~\ref{fig:personality} we report the Pearson correlation coefficients between personality traits and each job in form of radar plot. All the correlations are significant with a p-value smaller than $10^{-4}$. 

Among the 8 groups, people from start-ups are most open to new things, with a positive correlation of 0.13 to Openness. On the contrary, people of public relation are the most conservative. The correlation to Openness of them is -0.12. Software Engineers and marketing people are also open to new things, while office clerks and administrators people are the opposite. 

Marketing people, people of start-ups have higher conscientiousness levels. In other words, they show high motivations in their work. Office clerks have the lowest conscientiousness level. This agrees with our observation that office clerks post fewer content about their work. Moreover, software engineers, editors and writers, and designers also show a relatively low self-discipline.

Administrators are the most extroverted. A high positive correlation (0.18) to Extraversion indicates their strong preference for being in social situations. By contrast, the high negative correlation to Extraversion (-0.15) of software engineers shows low preference for social situations.

Office clerks are the most agreeable group of people. They have a positive correlation to Agreeableness of 0.20. Administrators also have a high Agreeableness level (0.18). Software engineers are the least agreeable. A negative correlation of -0.13 shows their low willingness to compromise.

The people from start-ups have highest stability level of emotion (-0.21 correlated to Neuroticism), followed by marketing people (-0.11). By contrast, a 0.24 correlation to Neuroticism of office clerks indicates they have relatively unstable emotions. Software Engineers and Designer also have positve correlations.

\section{Occupation Prediction}
\begin{figure}[!htbp]
\centering
\includegraphics[width= \columnwidth]{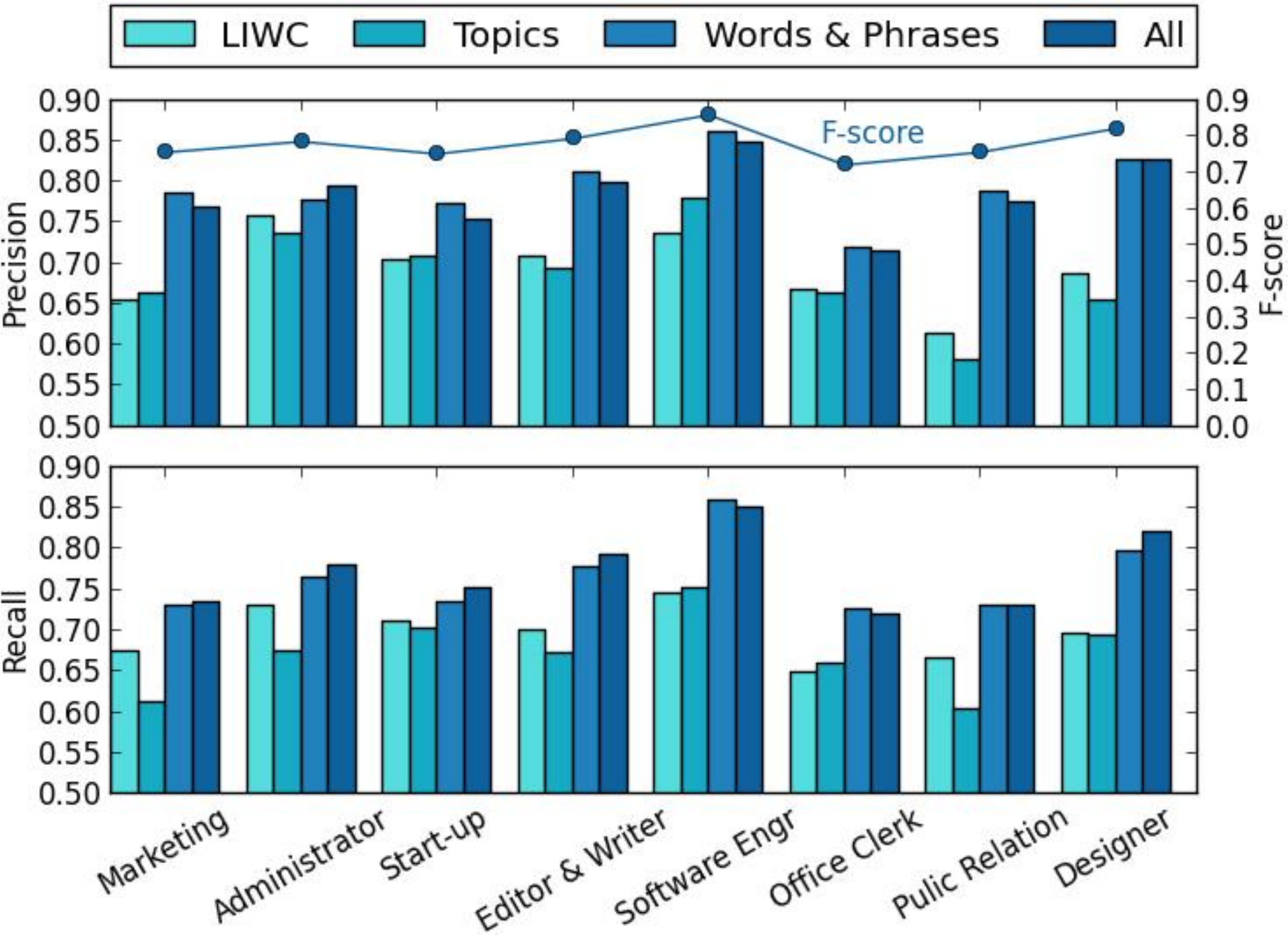}
\caption{Precisions and recalls of predicting 8 jobs using 4 sets of features. F-scores of using all features is also plotted.}~\label{fig:classification}
\end{figure}

Motivated by the observations on the correlation between human traits and occupations, we build a classifier to predict jobs based on features extracted from tweets. Each person is assigned a normalized weight on each job. We label the person with the job of the highest weight if the weight is larger than 0.8. If none of the weights is larger than 0.8, we remove the person from our data set. 
In each prediction task, we collect the people from a job as positive data, and then sample a equal number of people from other jobs as negative data. We tried 4 sets of features: LIWC features, 2,000 tweets topics, words and phrases, and all above. The values of words and phrases features are the TF-IDF values calculated for terms based on their frequencies in ones' tweets. The values of tweet topic features are the weights that each person is assigned on the topics. We plot the results in Figure~\ref{fig:classification}. The results are evaluated using 5-cross validation.

The average F-score of all eight jobs is 0.78, indicating the strong discrimination power of language features on job prediction tasks.

Software engineers (precision = 87\%, recall = 86\%) and Designers (precision = 82\%, recall = 83\%) have better prediction results. This is because software engineers and designers have more distinguishing language patterns and unique interests. People with both jobs talk much more about techniques, such as programming languages, online tools, while these technique terms are seldom used by people with other jobs. 
Because of the same reason, editors also have a relatively high precision (81\%) and recall (79\%). They mention a lot about reading and writing, and show stronger interest in politics and social events.
Office clerks have relatively low results (precision = 73\%, recall = 73\%). This is because although these group of people have unique preferences (strong self-awareness word like \textit{i}, \textit{my}, \textit{my life}), these features are also usually used by other people. 
Marketing, Administrator, Start-up, and Public Relation, have median performances, with precisions around 78\% and recalls around 75\%. This is because some language patterns and interests are shared among these four jobs. For example, people with these jobs are all interested in business, companies, and marketing. The prediction results align with the fact that Software Engineers and Designers are people whose skills are more specialized, whereas Marketing, Administrator, Start-up, and Public Relation require skills that are less specialized but more holistic, especially for the case of office clerk.

Words and phrases related features have the best performance among the three single type features. This is because words and phrases are the most informative type of features among three. They cover all the terms used in tweets with the cost of high dimensions of features. The lexicon LIWC uses is fixed and relatively small, while topic features are based on word clusters. Although these two types have fewer dimensions, the information is also reduced. When combining the three together, we observe a small boost of performance, especially on recall. This suggests that extra information brought by LIWC and topics is helpful the in classifying task.

\section{Conclusions and Future Work}
In this paper, we investigate the divergences across occupations. From multiple platforms, we gathered user information of several aspects. 
To overcome the ambiguity and uncertainty of self-reported information, a soft clustering approached was applied to extract occupations from crowdsourcing data. 
Linguistic styles were described using the most positively and negatively correlated words and phrases to people, while people's' interests are learned by extracting significant topics from their tweets. The Big Five Traits are also inferred using the tweet texts. We used Person Correlation Coefficients to uncover the differences of above human characteristics across jobs. The results indicate that people with different jobs have unique preferences to certain language styles and interests. Our results also reveal clear divergent levels of the Big Five Traits of different jobs. A classifier was built to predict people's job based on the features extracted from tweets. A high accuracy indicates the strong discrimination of language features on job prediction tasks. 

Overall, our study has revealed interesting patterns of how individuals' characteristics and daily tweets are connected to their job profiles. By showing the similarities and dissimilarities between job categories based on those characteristics and communication styles, our findings suggest that some individuals can possess set of characteristics that fit with multiple jobs, while others might possess a set of characteristics that are more unique to one specific type of job. 
In the future, we would like to introduce more features to categorize people into different job profiles, such as using their job titles and looking into their employment histories. We are also interested in extracting more human characteristics, besides personality and use of language, from their tweets; for examples, their life styles, habits, leisures. By obtaining richer information from people's online profiles, a more comprehensive study could be performed to uncover the deeper connection between people's personal lives and their jobs.

\section{Acknowledgment}
We would like to thank the support from the New York State through the Goergen Institute for Data Science, as well as Xerox Foundation.


\balance{}

\bibliographystyle{aaai}

\bibliography{reference}
\end{document}